\documentstyle[12pt]{article}
\begin{document}
\begin{titlepage}
\title{Wu-Yang Monopoles\\and\\Non-Abelian Seiberg-Witten Equations}
\author{ T. Dereli , M. Tekmen
\\{\small Department of Physics}\\{\small Middle East Technical University}\\
{\small 06531 Ankara, Turkey}}
\maketitle
\begin{abstract}
\noindent Some exact solutions of the 
SU(2) Seiberg-Witten equations in Minkowski spacetime
are given.
\end{abstract}
\end{titlepage}

\noindent {\bf 1. Introduction}
\vskip 2mm
The asymptotic freedom of non-Abelian gauge theories makes 
them unsuitable for doing reliable 
low energy calculations because in that limit
the theory enters into its strong coupling regime. 
Seiberg and Witten [1] used duality arguments within the framework
of N=2 supersymmetric Yang-Mills theory to bring an answer
to this long standing problem.
As a remarkable by-product Witten [2] has shown that
the Donaldson invariants of 4-manifolds can be determined
by essentially counting the solutions 
of a set of massless magnetic monopole equations of the
dual Abelian gauge theory [3],[4]. It was noted that the
Seiberg-Witten equations do not admit any square integrable solutions.
On the other hand Freund [5] has shown that  
the pair $(A,\psi)$ consisting of a Dirac monopole
potential together  with a simple ansatz 
describing a positive chirality spinor
field in Minkowski spacetime
satisfies the Seiberg-Witten equations
analytically continued to a Lorentzian spacetime. 
In fact the ansatz  $(A,\psi)$
was constructed by G\"{u}rsey [6] long before 
the Seiberg-Witten theory.
The simplicity of the Freund-G\"{u}rsey solution 
encourages the exploration of 
other solutions.
A non-Abelian generalization of Seiberg-Witten equations
have been recently considered both from the physical [7]
and  the mathematical [8] points of view.
The most natural SU(2) extension of  Seiberg-Witten
equations 
in Minkowski spacetime admits a Wu-Yang magnetic pole solution and
it is our purpose here to demonstrate this solution explicitly.
\vskip 4mm
\noindent {\bf 2. SU(2) Seiberg-Witten Equations}
\vskip 3mm
SU(2) monopole equations consists of the pair of equations
\begin{equation}
\rho^{+}(F) = k {\Psi}^{\dagger} \gamma \wedge \gamma \Psi 
\end{equation}
\begin{equation}
{}^{\ast} \gamma \wedge  D_{A} \Psi =0
\end{equation}
where $A=(A^1,A^2,A^3)$ is a su(2) Lie algebra valued 
Yang-Mills potential 1-form.
We may also write $A=A^{a}T_{a}$ where $\{T_{a}\}$ are the
anti-hermitian generators of su(2) satisfying 
$[T_{a},T_{b}]= \epsilon_{abc} T_{c}$.
The corresponding Yang-Mills field (curvature)  2-form 
\begin{equation}
F=dA+e A \wedge A.
\end{equation} 
$\rho^+$ projects the self-dual part of $F$.
We introduced a gauge constant $e$ for later convenience.
Then the local gauge transformations are given by
\begin{equation}
A \rightarrow U A U^{-1} + \frac{1}{e} U dU^{-1}.
\end{equation}
$\Psi=(\psi^1,\psi^2,\psi^3)$ is a su(2) Lie algebra valued 
2-spinor  of positive chirality. The exterior covariant derivative 
of $\Psi$ is defined as
\begin{equation} 
D_{A} \Psi =d\Psi + e[A,\Psi]. 
\end{equation}
The constant $k$ in (1) can always be absorbed by a scaling of $\Psi$.

We fix the orientation of space-time by letting 
${}^{\ast}1= e^0 \wedge e^1 \wedge 
e^2 \wedge e^3$. \newline
${}^{\ast}$ denotes the Hodge map related to the metric
\begin{equation} 
g=-e^o \otimes e^0 + \vec{e} \otimes \vec{e}
\end{equation}
where $\{ e^a=( e^0,\vec{e})  \}$ is an orthonormal coframe.
We used in (1) and (2) the following $Cl(3,1)$ Clifford algebra basis  
\begin{equation}
\Gamma = \left ( \begin{array}{cc} 0&\gamma\\ \gamma^{\ast}&0 \end{array}
\right )
\end{equation}
where  $\gamma=\gamma_{a} e^a$ and
$\gamma_{a}=(I,{\vec{\sigma}})$. $I$ is the $2 \times 2$ 
identity matrix and ${\vec{\sigma}}$ are the standart Pauli matrices.

\vskip 4mm  
\noindent {\bf 3. Static, rotationally symmetric solutions}
\vskip 3mm
We will be considering static solutions
so that it is convenient to work with either local Cartesian
coordinates $(t,x,y,z)$ or spherical polar coordinates 
$(t,r,\theta,\phi)$. These are related through the  coordinate transformations
$$ x = r \sin{\theta} \cos{\phi}, \hskip 2mm 
y = r \sin{\theta} \sin{\phi}, \hskip 2mm 
z = r \cos{\theta}. $$
We assume the gauge potentials are independent of $t$
and Coulomb gauge is chosen so that  $A_0=0$. 
We take $e=\frac{1}{2}$ and adjust  $k$ in a suitable way.
With these assumptions, the SU(2) Seiberg-Witten equations (1) and (2)
reduce to the following coupled 3-dimensional equations:
\begin{eqnarray}
{}^{\ast}{\sigma} \wedge (d\psi^1 - \frac{1}{2} A^2 \psi^3 + 
\frac{1}{2} A^3 \psi^2)&=&0 \\
{}^{\ast}\sigma \wedge (d\psi^2 - \frac{1}{2} A^3 \psi^1 + 
\frac{1}{2} A^1 \psi^3)&=&0 \\
{}^{\ast}\sigma \wedge (d\psi^3 - \frac{1}{2} A^1 \psi^2 + 
\frac{1}{2} A^2 \psi^1)&=&0 
\end{eqnarray}
\begin{eqnarray}
F^1&=& \frac{1}{4}( {\psi^2}^{\dagger}\Sigma \psi^3-
{\psi^3}^{\dagger}\Sigma \psi^2)\\
F^2&=& \frac{1}{4}( {\psi^3}^{\dagger}\Sigma \psi^1-
{\psi^1}^{\dagger}\Sigma \psi^3)\\
F^3&=& \frac{1}{4}( {\psi^1}^{\dagger}\Sigma \psi^2-{\psi^2}^{\dagger}\Sigma \psi^1)
\end{eqnarray}
where $\Sigma = \sigma \wedge \sigma$. 
\vskip 3mm
\noindent i) Abelian solution:
\vskip 2mm
\noindent We let $\psi^1=\frac{\sqrt{2}}{ 2 r} (\xi+\eta)$, 
$\psi^2=\frac{i \sqrt{2}}{2 r} (-\xi+\eta)$ and $\psi^3=0$,
where 
$$\xi = \left ( \begin{array}{c} \cos{\frac{\theta}{2}}e^{-i\phi}\\
\sin{\frac{\theta}{2}} \end{array} \right )  =
\frac{1}{\sqrt{2r(r-z)}} 
\left (\begin{array}{c} x-iy\\r-z \end{array} \right )$$ 
and
$$\eta = \left ( \begin{array}{c} \sin{\frac{\theta}{2}}\\ 
-\cos{\frac{\theta}{2}}e^{i\phi}\\
 \end{array} \right )  =
\frac{1}{\sqrt{2r(r-z)}} 
\left (\begin{array}{c} r-z\\-(x+iy) \end{array} \right )$$
Then $A^1=0, A^2=0, A^3=-(1+\cos{\theta})d\phi$. 
This solution describes a magnetic monopole of strength 1, 
whereas the Freund-G\"{u}rsey solution describes
a monopole of strength $\frac{1}{2}$.
We note a Dirac string singularity along the +ve z-axis [9].
\vskip 3mm
\noindent ii) Non-Abelian solution:
\vskip 2mm
\noindent We start with
Wu-Yang monopole potentials written in terms of Cartesian coordinates [10]
\begin{eqnarray}
A^1&=&\frac{1}{r^2}(zdy-ydz) \nonumber \\
A^2&=&\frac{1}{r^2}(xdz-zdx) \nonumber \\
A^3&=&\frac{1}{r^2}(ydx-xdy).
\end{eqnarray}
It should be noted that these are free of any line singularity.
A full solution is obtained for
\begin{eqnarray}
\psi^1&=&\frac{\sqrt{3}}{2r}(a \xi + a^{\ast} \eta) \nonumber \\
\psi^2&=&\frac{\sqrt{3}}{2r}(b \xi + b^{\ast} \eta) \nonumber \\
\psi^3&=&\frac{\sqrt{3}}{2r}(c \xi + c^{\ast} \eta)
\end{eqnarray}
where $$ a = \frac{(r-z)}{2r}-\frac{(x+iy)^2}{2r(r-z)}  , 
b = i\frac{(r-z)}{2r}+i\frac{(x+iy)^2}{2r(r-z)} ,   c= \frac{(x+iy)}{r}.$$
\vskip 4mm
\noindent {\bf 4. Concluding Comments}
\vskip 3mm
In fact the non-Abelian solution (14) and (15) above may be related 
to the Abelian solution by a singular 
gauge transformation followed by  a scaling of the spinor field.
The required gauge transformation is
given in  $2 \times 2$ matrix notation for convenience as [11]
\begin{equation}
U = e^{-i{\sigma}^{3} \frac{\phi}{2}} e^{i {\sigma}^{2} \frac{\pi-\theta}{2}}
e^{i{\sigma}^{3} \frac{\phi}{2}}.
\end{equation}
Then the transformation rules for the components of $\Psi$ are found from
$$ \left ( \begin{array}{cc} \psi^3&\psi^1 -i \psi^2\\ \psi^1+i\psi^2&-\psi^3
\end{array} \right )  \rightarrow
U \left ( \begin{array}{cc} \psi^3&\psi^1 -i \psi^2\\ \psi^1+i\psi^2&-\psi^3
\end{array} \right ) U^{-1} $$ by substituting $U$ from above. 

The Abelian solution can be obtained in any gauge group; for example $SU(N)$.
In this case the Lie algebra can be split into the Cartan subalgebra ${\bf t}$
and the roots $\alpha$, and we can set  all the gauge potentials in the
root directions to zero while setting all the monopole fields in the 
Cartan directions to zero. The Dirac equation for the monopoles becomes
$$  {}^{\ast}\sigma \wedge ( d \psi^{\alpha} + \alpha(A) \psi^{\alpha}) = 0.$$
This is solved by simply taking the Freund-G\"{u}rsey solution for the
monopoles and taking $\alpha(A) = A^0$ where $A^0$ is the Abelian Dirac
potential. For $SU(2)$ this implies that $A = 2  A^0$ which is why the 
magnetic charge  of our solution is 1 rather than $\frac{1}{2}$.
To complete the analysis, all that is left  is to work out
the gauge field strength 2-forms from the remaining equations.
This can be done consistently after an appropriate scaling of the spinors.
Once the Abelian solution to the $SU(N)$ Seiberg-Witten equations is 
found, it may then be gauge rotated to any non-Abelian solution.

It is to be expected that the moduli space of  
non-Abelian Seiberg-Witten monopole equations
has a structure richer than that of ordinary Donaldson theory.
Accordingly we have been able to show that SU(2) Seiberg-Witten equations
admit a simple Wu-Yang magnetic pole solution . 
We think this provides  one other reason why
SU(2) Seiberg-Witten equations should be investigated more closely.

\vskip 2cm
\noindent {\bf Acknowledgements}
\vskip 3mm
We thank the referee for the inclusion of
the SU(N) generalisation
 of the Abelian solution.

\newpage
\noindent {\bf References}
\vskip 3mm
\begin{description}
\item[[1]] N. Seiberg, E. Witten, Nucl.Phys. {\bf B426}(1994)19
\\(Err. ibid,{\bf B430}(1994)485)
\item[[2]] E. Witten, Math. Res. Lett. {\bf 1}(1995)1
\item[[3]] R.Flume, L.O'Raifeartaigh,I.Sachs, {\it Brief resume of
Seiberg-Witten theory} hep-th/9611118
\item[[4]] S. Akbulut, {\it Lectures on Seiberg-Witten invariants}
in{\bf Proceedings of 4th G\"{o}kova Geometry-Topology Conference} p.95
Edited by S.Akbulut, T.\"{O}nder, R.J.Stern (TUBITAK, Ankara,1996) 
\item[[5]] P.G.O.Freund, J.Math.Phys.{\bf 36}(1995)2673
\item[[6]] F. G\"{u}rsey in  {\bf Gauge Theories and Modern Field Theory}
Edited by R. Arnowitt, P. Nath (MIT Press, 1976)
\item[[7]] J.M.F.Labastida, M.Mari\~{n}o,Nucl.Phys.{\bf B448}(1995)373
\item[[8]] C.Okonek,A.Teleman,Comm.Math.Phys.{\bf 180}(1996)36
\item[[9]] P. A. M. Dirac, Proc. Roy. Soc. {\bf A133}(1931)60
\item[[10]] T.T.Wu,C.N.Yang, in {\bf The Properties of Matter Under Unusual
Conditions} Edited by H.Mark,S.Fernbach (Wiley,1969)
\item[[11]] J. Arafune, P.G.O.Freund, C.J.Goebel, 
J.Math.Phys.{\bf 16}(1975)433
\end{description}
\end{document}